\documentclass[conference]{IEEEtran}
\IEEEoverridecommandlockouts
\usepackage{cite}
\usepackage{amsmath,amssymb,amsfonts}
\usepackage{algorithmic}
\usepackage{graphicx}
\usepackage{textcomp}
\usepackage{xcolor}
\usepackage{hyperref}

\usepackage{siunitx}
\hypersetup{
    colorlinks=true,
    linkcolor=blue,
    filecolor=magenta,      
    urlcolor=blue,
    }
\def\BibTeX{{\rm B\kern-.05em{\sc i\kern-.025em b}\kern-.08em
    T\kern-.1667em\lower.7ex\hbox{E}\kern-.125emX}}

\makeatletter
\newcommand{\linebreakand}{%
  \end{@IEEEauthorhalign}
  \hfill\mbox{}\par
  \mbox{}\hfill\begin{@IEEEauthorhalign}
}
\makeatother

\begin{document}

\title{LIPSFUS: A neuromorphic dataset for audio-visual sensory fusion of lip reading \\
\thanks{This research was partially supported by the Spanish grants MINDROB (PID2019-105556GB-C33/AEI/10.13039/501100011033) and SMALL (PCI2019-111841-2/AEI/10.1309/501100011033) projects. E. P.-F. was supported by a "Formaci\'{o}n de Personal Universitario" Scholarship from the Spanish Ministry of Education, Culture and Sport. Authors thank the Ulster University people and their HPC. A. L-B thanks the Salvador de Madariaga mobility program.}
}

 \author{
 \IEEEauthorblockN{A. Rios-Navarro\textsuperscript{1}, E. Piñero-Fuentes\textsuperscript{1}, S. Canas-Moreno\textsuperscript{1}, A. Javed\textsuperscript{2}, J. Harkin\textsuperscript{2} and A. Linares-Barranco\textsuperscript{1}}
 \IEEEauthorblockA{\textsuperscript{1}\textit{Robotics and Technology of Computers Lab. I3US. SCORE. University of Seville}. Seville, Spain }
 \IEEEauthorblockA{\textsuperscript{2} \textit{School of Computing, Eng \& Intel. Sys.  Magee campus.} Derry, UK}
 \IEEEauthorblockA{arios@us.es}
}
\maketitle

\begin{abstract}
This paper presents a sensory fusion neuromorphic dataset collected with precise temporal synchronization using a set of Address-Event-Representation sensors and tools. The target application is the lip reading of several keywords for different machine learning applications, such as digits, robotic commands, and auxiliary rich phonetic short words. The dataset is enlarged with a spiking version of an audio-visual lip reading dataset collected with frame-based cameras. LIPSFUS is publicly available and it has been validated with a deep learning architecture for audio and visual classification. It is intended for sensory fusion architectures based on both artificial and spiking neural network algorithms.
\end{abstract}

\begin{IEEEkeywords}
Neuromorphic dataset, sensory fusion, dynamic vision sensor, neuromorphic auditory sensor
\end{IEEEkeywords}

\section{Introduction}
Sensor fusion is known as the process of combining sensor data derived from several sources of the same reality such that the fused information has less uncertainty than when these sources are used individually. In \cite{radar_vision_fus_2007_TITS} authors demonstrated that combining radar and visual information improves the accuracy in vehicle detection systems, where radar information is used to focus on the important part of the visual information. In the healthcare field sensory fusion improves decision-making when combining data from different sensors, like in \cite{health_senfus2021}, where up to eight data sources are combined for the detection of diabetes. In urban search and rescue robots the sensory fusion goes further and combines proprioceptive (inertial measurement unit and tracks odometry) and exteroceptive sensors (omnidirectional camera and rotating laser rangefinder) to improve the accuracy \cite{multimodal_sens_robots2015}. In \cite{sensfus_quality2022} a review of sensory fusion for quality control in manufacturing is presented, where different kinds of sensors are combined including visual, acoustic, laser, vibration, thermal, etc. In general, data fusion is a challenging task because of a several difficulties. The majority of these difficulties arise from the data being fused, imperfection and diversity of the sensor technologies, and the nature of the application environment as data imperfection, outliers and spurious, conflicting, modality, correlation, alignment, association data, operational timing, etc. \cite{dataFus_review_2013}.

Living organisms usually have several sensory mechanism to interact with the real world. Those with neural architectures, learn from the experience taken from sensory systems and their combination or fusion. Neuromorphic engineering is devoted to study these and others neural systems in biology by the implementation of engineering systems that mimics those present in biology \cite{NE_Mead_1990}. One of those neural architectures is in charge of audio-visual sensory fusion, where the temporal synchrony is one of the strongest binding cues in multi-sensory perception \cite{multisensoryIntegration_2005}\cite{multisensoryIntegration_2013}, where the consideration of a temporal window is of utmost importance for this type of sensory fusion \cite{windowTempIntegration2007}. This window temporal length decreases in humans with the age \cite{audiovisualTempFus2014}. 

Audio-visual sensory fusion developments with neuromorphic engineering open the range of applications i.e, Mobile robotics, Internet of Things (IoT), Edge Computing, etc where low latency and low power consumption are important factors. The use of spiking sensors and spiking neural networks will considerably improve these characteristics over conventional sensors and artificial neural networks. 

This paper is focused on the collection of a neuromorphic audio-visual dataset for machine learning applications around lip reading. It takes into account the temporal synchronization of the measured data using specific neuromorphic sensors and hardware tools. The dataset is collected from people of different nationalities and age, speaking the number of Natural Language Processing (NLP) words and one English-language pangram \textit{"The quick brown fox jumps over the lazy dog"} for training of learning architectures. The dataset is publicly available and it has been tested, for validation, with convolutional neural networks. 
To the best of our knowledge, there is no previous neuromorphic dataset recorded with spiking sensors and specific logic to ensure temporal synchronization. 

\section{Materials and methods}
This section explains the setup for the dataset collection that has been recorded directly from neuromorphic sensors while maintaining the time synchronization of the visual and audio parts. It also explains how the BBC dataset has been converted to the spikes domain. 

\subsection{LIPSFUS recording setup}
The Neuromorphic Auditory Sensor (NAS) \cite{nas_angel17} and the Neuromorphic Dynamic Vision Sensor (DVS) \cite{cnmdv_Teresa13} has been used to record the LIPSFUS dataset. Both sensors generate spike information at their outputs and both are encoded in AER (Address-Event-Representation) format at their outputs using a digital parallel bus (16-bit maximum in our setup) with a 2-bit asynchronous handshake. The digital words encode the identifier of the emitter neuron, in the corresponding sensor, called the address. The length of an address is fixed by the number of neurons in the sensor: pixels for the DVS or channels for the NAS. In our setup, it is used the DVS developed at IMSE-CNM in Seville, called cnmDV. This DVS has 128x128 pixels, so the output parallel AER bus requires 15-bit for the addresses plus the polarity bit. For the audition, the selected NAS has been obtained from the openNAS tool \cite{gutierrez2021opennas} and synthesized for the Spartan6 FPGA of the AER-Node board \cite{aernode_tbiocas17}. A binaural sensor composed of two identical banks of 64 band-pass filter per ear has been selected, with the cut-off frequencies shown in figure \ref{fig:NAS-freq} in the range from 18,91Hz to 20,81KHz, with an average error of 0,005\% and a standard deviation of 0,001 along the 64 channels. For this audio sensor, the AER bus is sending addresses of 8-bit length (6-bit to identify an active channel, 1-bit for polarity and 1-bit for left or right filter bank). The input of the NAS comes from a stereo set of ear-shaped microphones that mimics a human head. This is the 3DIO Free Space XLR Binaural setup \cite{microphones}.
These two neuromorphic sensors produce spikes with their own temporal distributions, that intrinsically depend on the sensed activity. An AER-tool \cite{aertools} is used for merging the two sensors' activity in such a way that the temporal distribution is maintained for both sensors. This requires the AER-merger to add 1-bit into the AER bus to distinguish between a visual spike or an audio spike. The most significant bit is set to '0' for a DVS event, and to '1' for a NAS event. Therefore, the full range of 16-bit allowed by the hardware is used in this setup. The output of the AER-merger is then connected to an AER-monitor board \cite{usbaermini2}, that assigns a timestamp mark to each received event, regardless of which sensor it belongs to, before queuing the data (address and timestamp) for sending USB packets to a computer running jAER, which stores .AEDAT files for each sample of the dataset.  


\begin{figure}[h!]
 \centering
 \includegraphics[scale=0.5]{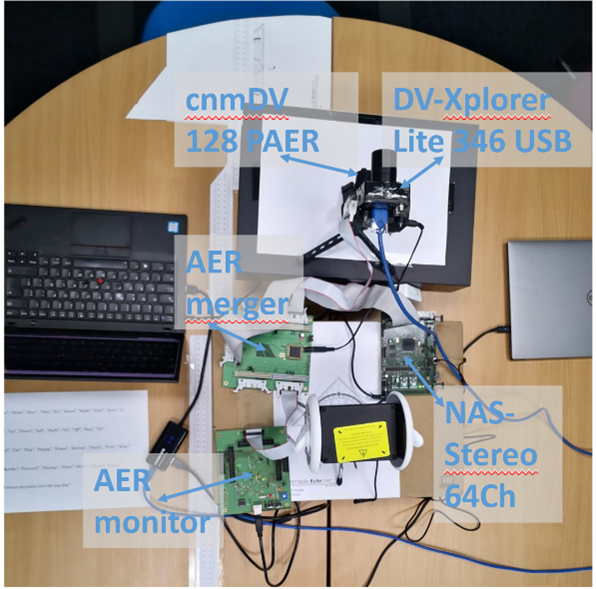}
 \includegraphics[scale=0.53]{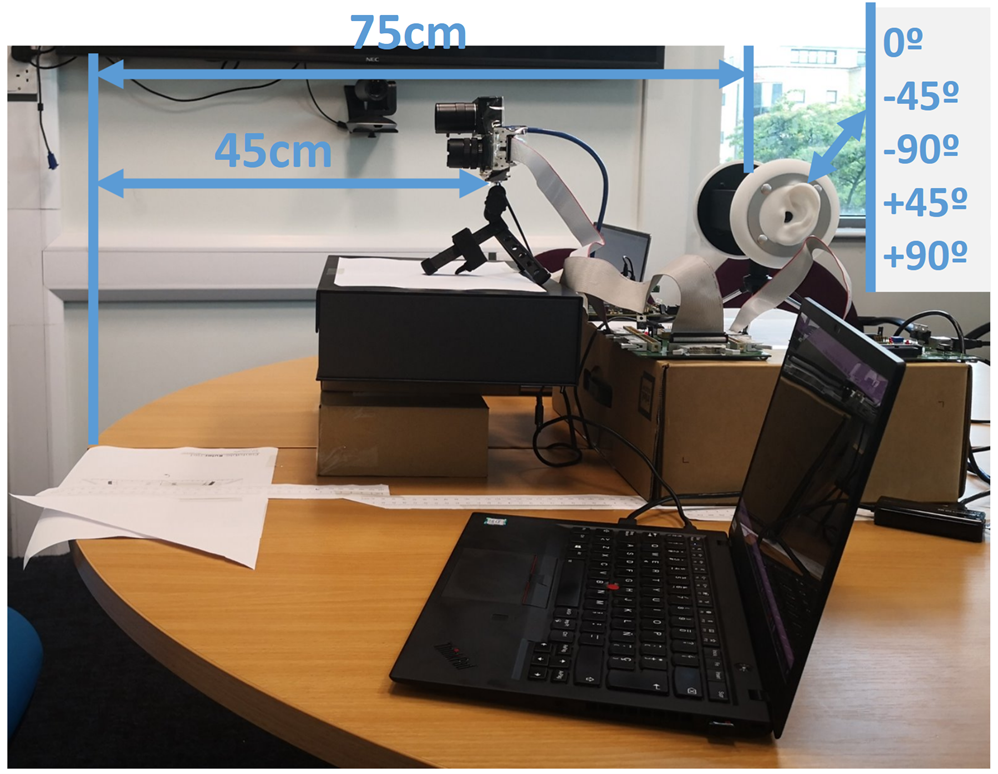}
 \caption{LIPSFUS recording dataset setup. Left figure shows the sensors and boards distribution and the right figure shows the setup sensor distances.}
  \label{fig:setup}
\end{figure}

\begin{figure}[h!]
 \centering
 \includegraphics[width=1.00\linewidth]{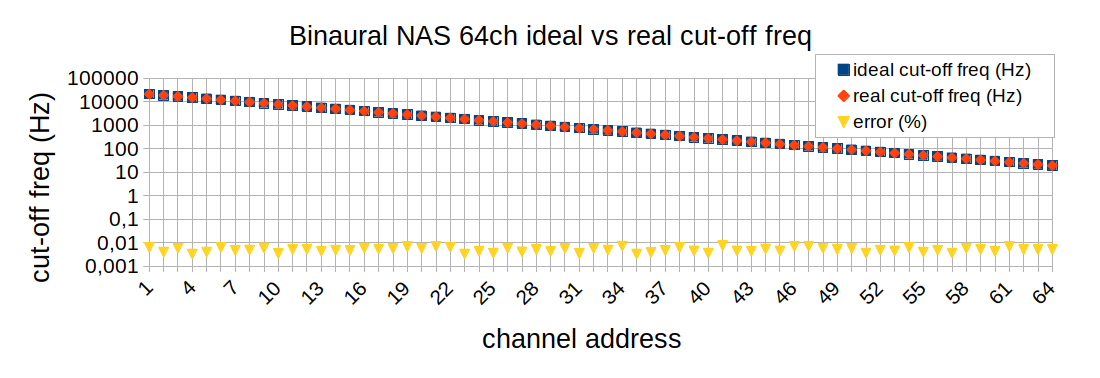}
 \caption{Cut-off frequencies of the binaural NAS 64-channels. Ideals versus real implemented ones on FPGA with error in \%.}
  \label{fig:NAS-freq}
\end{figure}

Figure \ref{fig:setup} on the left shows the distribution of the neuromorphic sensors (cnmDV and NAS), the AER-merger platform and the AER-monitor. In addition, another neuromorphic sensor, DV-Xplorer-Lite-346-USB, can be seen in the figure, but it has not been included in the dataset presented in this paper. It is incorporated in the setup for future use in applications that requires higher visual resolutions. The right side of the figure shows the distances of each neuromorphic sensor to the person to be recorded. The cnmDV is placed at 45cm from the person face, while the 3DIO microphones are placed at 75cm and they can be oriented at different angles (0º, -45º, -90º, 45º and 90º in our dataset). The microphones are placed in the holes of two silicone ears that simulate the human ear, and both are separated at the same distance as in an average human head.
The dataset is recorded in two different environments: noisy and quiet. The noisy environment consisted of a glass meeting room, where the air-conditioning system was running, and behind the window was a car park and a main street in the city. In contrast, the quiet environment consisted of a small room with acoustically insulated walls and a door. The lighting condition in both environments were kept similar. The dataset, for each of the environments, consisted of 22 persons of 5 different nationalities (Indian, Iranian, Irish, Pakistani, Spanish), of both genders and aged between 6 and 61 years.

The dataset consists of a series of words that have been selected from different challenges or are considered to be of interest in the area of language processing. The words in the dataset are:

\begin{itemize}
  \item Spoken digits: One, two, three, four, five, six, seven, eight, nine, zero and o \cite{warden2018speech}.
  \item Robotic commands: Yes, no, up, down, left, right, on, off, stop and go \cite{TensorFlow_Speech_Commands}.
  \item Bed, bird, cat, dog, happy, house, Marvin, Sheila, tree, wow \cite{TensorFlow_Speech_Recognition_Challenge}.
  \item About, border, forward, missing, press, short, threat, young \cite{chung2016lip}.
  \item The quick brown fox jumps over the lazy dog \cite{yang2019wide}.
\end{itemize}

Each participant was reading each word (or sentence) when it is shown in a presentation with 2 seconds delay between words. An AEDAT file was recorded for the whole presentation from the word "one" to the "fox" sentence. This is repeated five times per participant placing the ears at the orientations commented before. Therefore, a total of 5 different AEDAT recordings are stored for the each person. A repository with access to these files is available on GitHub\cite{LIPFUS_dataset}.

\subsection{BBC-LIPSFUS dataset}

The forth set of words recorded with the previous setup are also part of the "Lips reading in the wild" dataset from BBC \cite{chung2016lip, Chung17, Chung17a}. In this work, this set of words has been converted to a neuromorphic format and made available for training spiking learning architectures.
This dataset includes 1k sentences with 500 different words taken from news and interviews (MP4 videos) from BBC channels with 1k different speakers\cite{BBC_dataset}. Each MP4 video has 29 frames (1.16 seconds) and selected word is in its temporal center.
An important difference from our recorded dataset is that our speakers pronounced each word / sentence in an isolated way. Nevertheless, for the BBC dataset, the words are part of a sentence and they are pronounced in a natural conversation. In order to properly extract the required word from each MP4, we have used the IBM Watson Studio engine \cite{ibmwatson}, as it was used in \cite{chung2016lip} for validation, to extract the precise timestamps where our chosen words start and end at each MP4. Using these timestamps, each MP4 was cut to contain only the right word, in an isolated way, as it is in our recorded dataset. These cut MP4s were then used for the neuromorphic conversion.
The same neuromorphic auditory sensor (NAS) used in the previous recordings was connected to the audio output of a computer and used to capture .AEDAT files in jAER while the computer was playing each MP4 file of the dataset for the 8 selected words and 1k speakers of each word.
For the visual neuromorphic representation of the MP4 videos, the ESIM \cite{esim_pmlr18} has been used, which also produces AEDAT files for each of the 8k given inputs. ESIM behaves as the DVS for the presented input video.

\subsection{Dataset validation}
This dataset has been validated by performing a classification task on a subset of the words in the dataset, the spoken digits. This paper proposes an Artificial Neural Network (ANN) based learning model to classify extracted neuromorphic words dataset. Currently, the authors are also working on the design of a Spiking Neural Network (SNN) to perform the classification task using the spiking information from the sensors.

\subsubsection{Data conversion and augmentation}
This work has considered Convolutional Neural Network (CNN), to perform the classification of the spoken digits, the first step is to convert the spiking information into a data type that the CNN can process i.e, an histogram based image. As we have information from two different sensors that generate different nature of temporal information. Therefore a separate conversion is designed for each sensory data.

\begin{figure}[h!]
 \centering
 \includegraphics[width=0.8\linewidth]{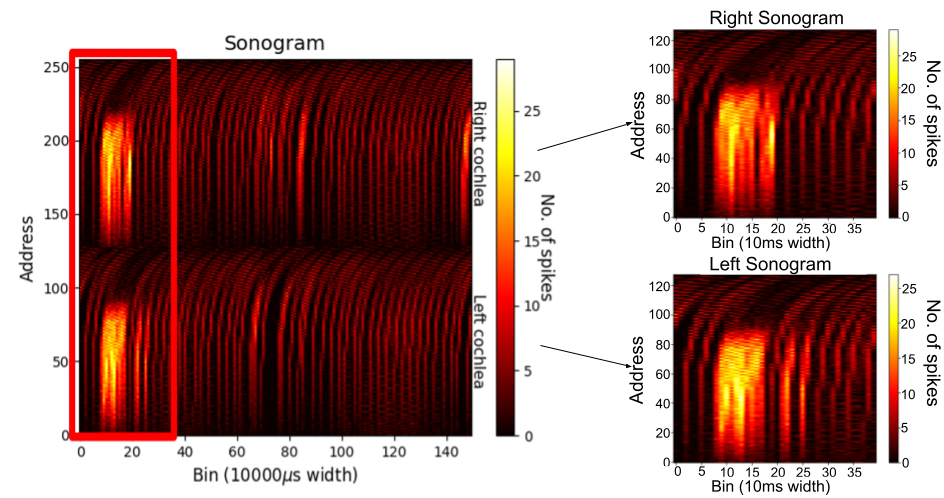}
 \caption{Stereo sonogram of a spoken digit sample (left). The spoken word information is delimited by the red box. Mono sonogram for right (top-right) and left (bottom right) channels cut from the red box.}
  \label{fig:stero_sonogram}
\end{figure}

The NAS used for the recording of this dataset consists of 64 stereo channels. The number of channels corresponds to the number of spike-based filters used to decompose the signal into different frequencies. As this sensor is stereo, it has 128 channels in total. At the output of each of these spike-based filters there are two neurons, one encoding the positive part of the signal and one encoding the negative part of the same signal. Therefore, this NAS contains 256 neurons that emit information at its outputs.
For the conversion of the pulsed information of the NAS, we proceeded to generate the sonogram of each spoken digit sample. The sonogram shows the activity of each sensor channel over time. To calculate the activity of each channel, a time window is used in which the number of spikes emitted by each channel is accumulated and grouped and the channel value is set to that accumulation for that time window. Thus, the information of each channel is encoded in the intensity of spikes produced in each time window. The figure \ref{fig:stero_sonogram} (left) shows the sonogram of a sample where on the x-axis is the time, on the y-axis are the neurons of the channels and the colour represents the intensity of the channel.

As can be seen in figure 3, the spoken word information is at the beginning of the recording (delimited by the red rectangle). It can be seen that there is a large part of the recording with some activity in the channels, which corresponds to some noise. Part of this noise is from the environment as well as noise generated by the sensor at the output of the filters. This work focused on the part of the recording that contains the most activity, the part delimited by the red rectangle, and we will also divide the stereo recording to obtain two mono samples as shown in the figure \ref{fig:stero_sonogram} (right).


In order to obtain more samples from the recorded dataset, a data augmentation process has been carried out. This is done by starting with a temporal window that is adjusted to the region where the spoken word activity is located and enlarging that region by allowing some noise prior to the spoken word activity. Then, to generate new samples, this region of noise prior to the activity is reduced a little and enlarged by the same amount at the end of the spoken word activity allowing for some noise. This process was repeated 10 times per sample. The Number of Samples (NS) in the audio dataset is 26,620 (see equation \ref{equ:audio_dataset_samples}, for 22 People (P), 11 Words (W), 2 channel audio Sensor (S), 10 Data Augmentation (DA) techniques and 5 different microphones Orientation (O).

\begin{equation}
\begin{aligned}
    NS = P \times (W \times S + W \times S \times DA) \times O \\
    = 22 \times (11 \times 2 + 11 \times 2 \times 10) \times 5
    \label{equ:audio_dataset_samples}
\end{aligned}
\end{equation}

\begin{figure}[h!]
 \centering
    \includegraphics[width=0.6\linewidth]{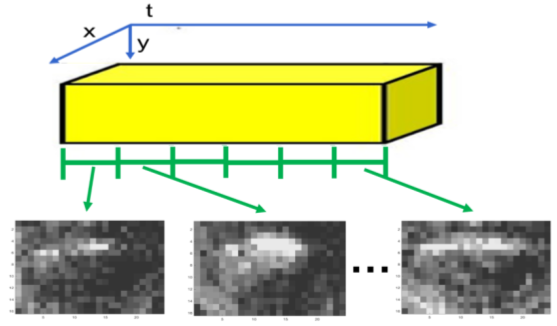}
 \caption{Conversion process from spiking information to histograms.}
  \label{fig:dvs_data_converion}
\end{figure}

The DVS used to record this dataset has a resolution of 128x128 neurons that emit positive and negative spikes depending on the change in brightness that the scene represents. For the conversion of the spiking information from the DVS sensor, only the lip area of the original sample was trimmed, discarding the rest of the information from the scene. The entire recording is then divided into equal temporal regions and the spikes are accumulated to generate a histogram equivalent to an image. The activity of each pixel or neuron in the sensor is encoded in the colour intensity of the pixel in the generated image. This approach provides a sequence of images that capture the movement of the lips when saying the corresponding word. Figure \ref{fig:dvs_data_converion} shows how this conversion is performed.

A similar data augmentation process has been repeated for the audio dataset to generate more samples. A temporal window is adjusted to the region where the lip movement activity is found when the word is spoken. This region is enlarged thus allowing some noise before the activity of the spoken word. To generate new samples, the selected region of noise prior to the activity is reduced marginally and enlarged by the same amount at the end of the speech activity allowing for some noise. This process was repeated 10 times per sample. The number of samples in the visual dataset is 13,310 (see equation \ref{equ:visual_dataset_samples}):

\begin{equation}
\begin{aligned}
    NS = P \times (W + W \times DA) \times O \\
    = 22 \times (11 + 11 \times 10) \times 5
    \label{equ:visual_dataset_samples}
\end{aligned}
\end{equation}


\subsubsection{Visual and Audio networks}
As mentioned at the end of the introduction section, to validate the dataset, two CNNs have been used to classify the spoken digits, one for the audio part and one for the visual part. The aim is to demonstrate with a simple application that the data collected in this dataset are valid for use by the community.
The CNN architecture that will be trained to classify the spoken digits using the audio information from the sonograms (via NAS sensor) has the following layout:
\begin{itemize}
    \item Input(64,32,1) $\,\to\,$ Conv2D(3,3,100) $\,\to\,$ MaxPool(2, 2) $\,\to\,$ Conv2D(2,2,200) $\,\to\,$ MaxPool(2,2) $\,\to\,$ Dropout(0.5) $\,\to\,$ Dense(10).
\end{itemize}
In contrast, the CNN architecture that will be trained to classify the spoken digits using the visual information from the sequence of histograms (via the DVS sensor) has the following configuration:
\begin{itemize}
    \item Input(33,49,6) $\,\to\,$ Conv2D(3,3,16) $\,\to\,$ MaxPool(2,2) $\,\to\,$ Conv2D(2,2,32) $\,\to\,$ MaxPool(2,2,2) $\,\to\,$ Conv2D(2,2,64) $\,\to\,$ MaxPool(2,2,2) $\,\to\,$ Dropout(0.5) $\,\to\,$ Dense(10).
\end{itemize}
Both networks have been trained using the Keras+Tensorflow framework.

\section{Results}
The Keras+Tensorflow framework was used to train the CNNs described in the previous section. For the different training sessions performed, the datasets have been partitioned into 70\% for training, 20\% for validation and the remaining 10\% has been used to test the network once it has been trained. The samples of the 10\% used for testing belong to a subject whose samples have not been used in the training phase, thus ensuring that during the test all the samples are totally unknown to the trained network.

Table I shows the results obtained when testing the audio (A\_) and the visual (V\_) CNNs across different training scenarios. Each row of the table represents the result of testing the CNN using the specified number of training epochs (Tr\_ep) and batch size (B\_size). The learning ratio has been set to the same value for all training epochs, in this case to 0.001.

\begin{table}[!h]
\centering
\begin{tabular}{|l|l|l|l|l|}
\hline
Tr\_ep; B\_size & A\_Loss & A\_Acc & V\_Loss & V\_Acc                                                                             \\ \hline
100; 32                       & 0.661             & 0.853            & 1.266              & 0.685                                 \\ \hline
200; 32                       & 0.704             & 0.882            & 1.191              & 0.694                                 \\ \hline
100; 64                       & 0.589             & 0.850            & 1.030              & 0.709                                 \\ \hline
200; 64                       & \textbf{0.475}    & \textbf{0.896}   & \textbf{0.997}     & \textbf{0.727}                        \\ \hline
100; 128                      & 0.605             & 0.832            & 1.099              & 0.701                                 \\ \hline
200; 128                      & 0.591             & 0.824            & 1.145              & 0.698                                 \\ \hline
\end{tabular}
\label{tb:audioVisual_results}
\caption{Test audio and visual results using different training parameters. The learning rate value is 0.001 for all.}
\end{table}

In the table the best test loss and accuracy values are highlighted for both cases. These results do not seem promising initially when compared to other applications that can be found in the literature, however the aim of this work is not to design a network model that is optimal in the classification of the spoken digits, but rather to validate the dataset in order to make it available to the community. 

\section{Conclusion}
This paper presents a dataset in which there is visual and auditory information when speaking a set of words. The visual information consists of the lip movement when the subject articulates a word, while the auditory information refers to the sound that the subject makes when pronouncing a word. This information has been captured by the NAS sensor for the audio and the DVS sensor for the visual part, synchronising the information from both sensors with the same timing source, therefore, achieving a perfect temporal sequencing of the spikes generated by both sensors. To validate the dataset, a classification task has been performed on a subset of words from the dataset using Deep Learning algorithms. Although the reported classification results are not the same as those obtained with Deep Learning techniques in the literature, they are satisfactory to validate that the dataset samples are useful and suitable for the research community to implement sensory fusion algorithms.

\bibliographystyle{IEEEtran}
\bibliography{IEEEabrv,IEEEexample}
\vspace{12pt}

\end{document}